\documentclass[fleqn,10pt]{wlscirep}
\usepackage[utf8]{inputenc}
\usepackage[T1]{fontenc}
\usepackage{lineno}
\usepackage{longtable}
\usepackage{multirow}

\title{A multi-sensor human gait dataset captured through an optical system and inertial measurement units}

\author[1,*]{Geise Santos}
\author[2]{Marcelo Wanderley}
\author[3]{Tiago Tavares}
\author[1]{Anderson Rocha}
\affil[1]{University of Campinas, Institute of Computing, Campinas, Brazil}
\affil[2]{McGill, Music Tech, Montreal, Canada}
\affil[3]{University of Campinas, School of Electrical and Computer Engineering, Campinas, Brazil}

\affil[*]{corresponding author(s): Geise Santos (geise.santos@ic.unicamp.br)}

\begin{abstract}

Different technologies can acquire data for gait analysis, such as optical systems and inertial measurement units (IMUs). Each technology has its drawbacks and advantages, fitting best to particular applications. The presented multi-sensor human gait dataset comprises synchronized inertial and optical motion data from 25 subjects free of lower-limb injuries, aged between 18 and 47 years. A smartphone and a custom micro-controlled device with an IMU were attached to one of the subject's legs to capture accelerometer data, and 42 reflexive markers were taped over the whole body to record three-dimensional trajectories. The trajectories and accelerations were simultaneously recorded and synchronized. Participants were instructed to walk on a straight-level walkway at their normal pace. Ten trials for each participant were recorded and pre-processed in each of two sessions, performed on different days. This dataset supports the comparison of gait parameters and properties of inertial and optical capture systems, whereas allows the study of gait characteristics specific for each system.
\end{abstract}
\begin{document}

\flushbottom
\maketitle

\thispagestyle{empty}

\section*{Background \& Summary}

Gait analysis has been explored since the 17\textsuperscript{th} century~\cite{baker2007history}. Advances in understanding the human motion throughout the last centuries allowed researchers apply this analysis to many applications, such as clinical assessment, monitoring of sports and athletic performances,  rehabilitation support, robotics research, and biometry-based recognition~\cite{muro2014gait}. This type of analysis can use data acquired by means of different technologies, like optical systems, inertial measurement units (IMUs), force-plate platforms, force shoes, and techniques based on computer vision. Although some of these are commonly used in most fields, as optical systems and imaging techniques, each technology has its drawbacks and fits best to particular applications~\cite{gouwanda2008emerging, winter2009biomechanics, morris2010review, ramirez2017review,yang2019review}.

One popular technology to acquire data for gait analysis is the optical motion capture system. It has minimal impact on the natural motion of the subject, as it does not need tethering any hardware onto the individual~\cite{kaufman1999future}. This system also fosters a precise acquisition of physical movements over virtual modeling and accurate reconstruction of movement marks and subjects' geometry~\cite{akhtaruzzaman2016gait}. However, optical motion capture systems are usually expensive, require high-speed processing devices and specific installations in a controlled space for their use~\cite{gouwanda2008emerging}.

Recently, IMUs have been considered an appropriate option to perform gait analysis because they mitigate these drawbacks of optical motion capture systems. They are typically more cost-effective, do not need a controlled environment, and support indoor and outdoor places. However, they have other limitations as being more susceptible to drift caused by changes in motion direction, and to low-frequency noise from small vibrations during the capture. Also, the sensor attachment position significantly impacts the estimation of gait parameters~\cite{akhtaruzzaman2016gait, mayagoitia2002accelerometer, gouwanda2008emerging, lee2010identifying}.

Several important gait datasets comprising either optical motion capture data or inertial data have been made available~\cite{moore2015elaborate, fukuchi2018public, schreiber2019multimodal, ngo2014largest, gadaleta2018idnet, luo2020database, santos2020manifold} in the prior literature. However, there are currently no datasets with data being simultaneous captured from both systems, which may allow a multi-modal gait analysis. This synchronized capture has proven helpful in specific applications, such as music gesture analysis~\cite{freire2020evaluation} and sports science~\cite{lapinski2019wide}. We propose, in this work, a multi-sensor gait dataset, which consists of inertial and optical motion data, and aims to provide basis for comparison and reasoning of human gait analysis using data from both systems.

The presented dataset comprises inertial and optical motion data from 25 subjects free of lower-limb injuries, aged between 18 and 47 years. A smartphone and a custom microcontroller device with an IMU were attached to one of the subject's legs to capture accelerometer data, and 42 reflexive markers were taped over the whole body to record three-dimensional trajectories. The participants were instructed to walk on a straight-level walkway at their normal pace. The custom device uses a  wireless protocol to communicate with the computer to which the optical system was connected. This setup enabled recording and synchronizing the trajectories (acquired by the optical system) and the accelerations (acquired by the dedicated device and the smartphone). Ten trials for each participant were recorded and pre-processed in each of two sessions, performed on different days. This amounts to 500 trials of three-dimensional trajectories, 500 trials of accelerations from the custom device, and 500 trials of accelerations from the smartphone. 

In addition to contributing with a multi-sensor dataset which supports the comparison of gait parameters and properties of inertial and optical capture systems, the full-body marker set and the inertial sensors attached to the leg favor the study of gait characteristics specific for each system. This dataset also allows analyzing gait variations between subjects and for each one (i.e., intra and inter-subjects) by the captures in different days. This characteristic of the dataset fosters investigations about the effectiveness of gait recognition and user profiling using accelerometers.

\section*{Methods}
\label{sec:methods}

\subsection*{Participants}
Twenty-five subjects (12 women, 12 men, and one undeclared gender, aged between 18 and 45 years) participated in this study, which took place between December of 2019 and February of 2020. Neither of them reported injuries for both legs or medical conditions that would affect their gait or posture. The participants were either students of the School of Music at McGill University, members of CIRMMT, or authors of this work. McGill University's Research Ethics Board Office approved this study (REB File \# 198-1019), and all participants provided informed consent.

\subsection*{Experimental design}
\begin{figure}[ht]
\centering
\includegraphics[width=0.9\linewidth]{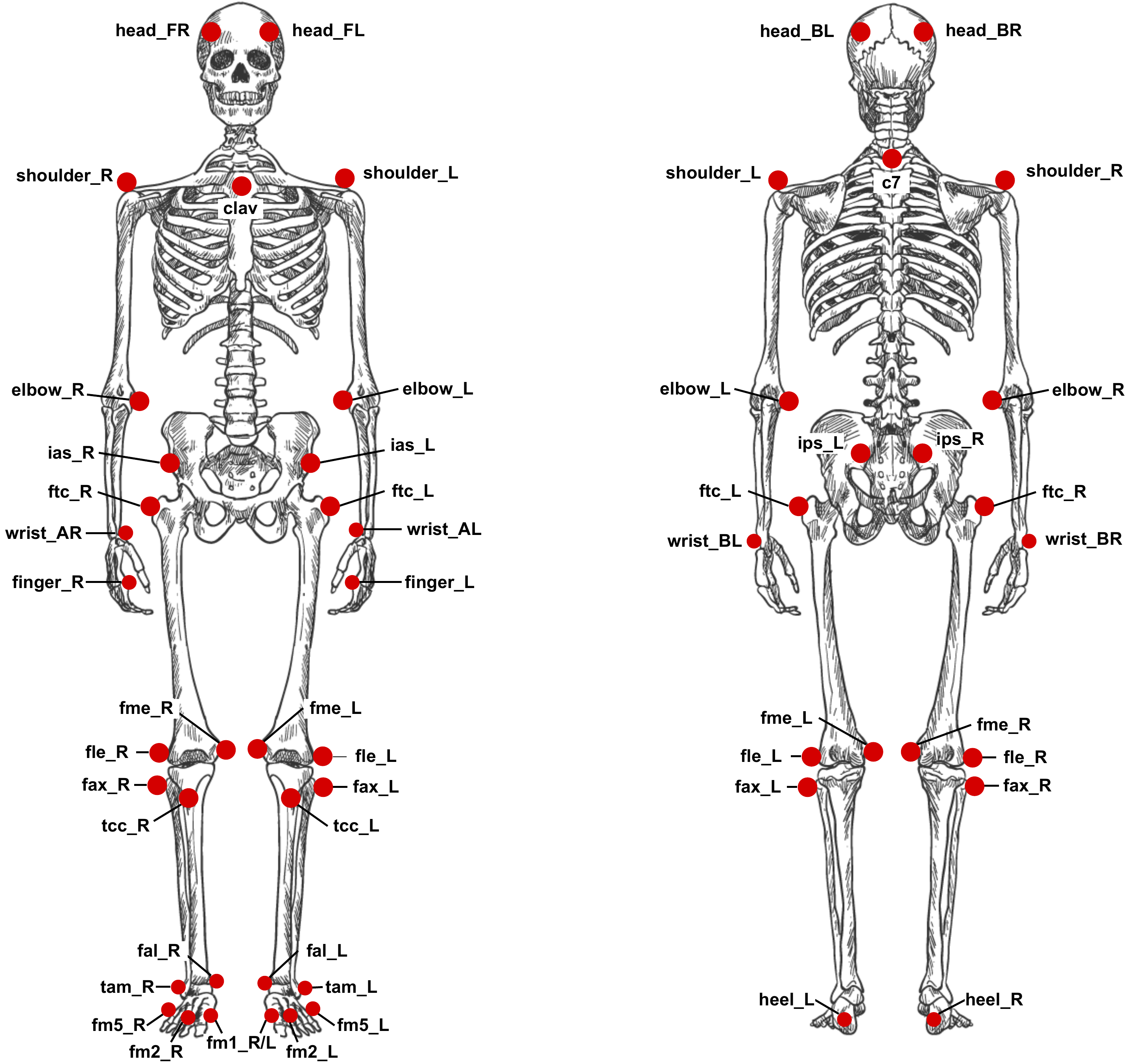}
\caption{The adopted marker set based in the model proposed by Dumas and Wojtusch and the Plug-in Gait model.}
\label{fig:markerset}
\end{figure}

A proprietary optical system~\cite{qualisys2021}, with 18 infra-red Oqus 400 and Oqus 700 cameras and sampled at 100Hz, was adopted to track the 3D trajectories of 42 reflective markers over the participants' body. The marker set was based on the lower-limb IORGait model proposed by Leardini et al.~\cite{leardini2007new}, and a simplified upper-limb and trunk Plug-in Gait models~\cite{davis1991gait}. This marker set is depicted in \autoref{fig:markerset}, and each reflexive marker, as well as its anatomical landmarks, are described in \autoref{tab:markers_details}. Gait analysis usually focuses on the lower limb trajectories; thus, the upper-limb and trunk simplified models are used only for skeleton reconstruction purposes. The marker placement was performed by anatomical palpation using the landmarks reported in \autoref{tab:markers_details} at the beginning of each session, and was not changed during the session trials. The proprietary motion capture software (\textit{Qualisys Track Manager} - QTM)~\cite{qtm2011} was employed to record and pre-process the 3D trajectories of the reflexive markers. All trajectory measurements were acquired in units of millimeters.

A Nexus 5 Android-based smartphone with an InvenSense MPU-6515 six-axis IMU was used to capture accelerometer data. An Android application was designed and developed to read accelerations at a sampling rate of 100Hz and store them into a comma-separated values (CSV) file. The accelerometer measurements yielded by the Android platform are in units of $m/s^2$. 

\begin{figure}[ht]
\centering
\includegraphics[width=0.5\textwidth]{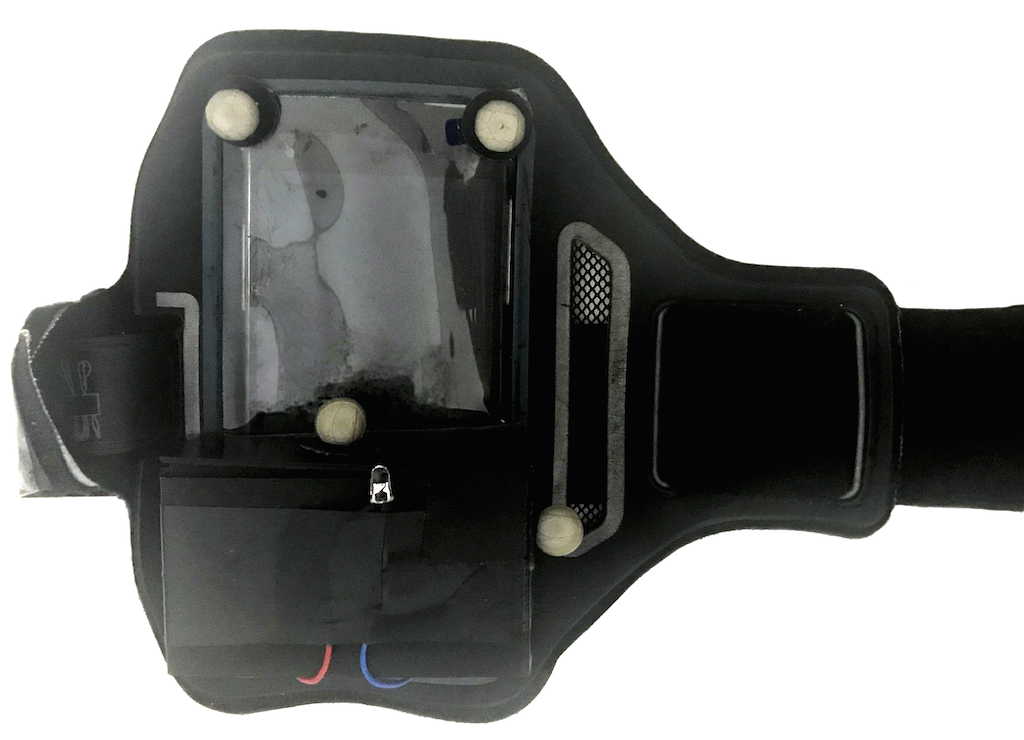}
\caption{The smartphone and MCU box mounted to a band to be attached to the leg.}
\label{fig:nexus_imu_band}
\end{figure}

\begin{figure}[ht]
\centering
\includegraphics[width=0.3\textwidth]{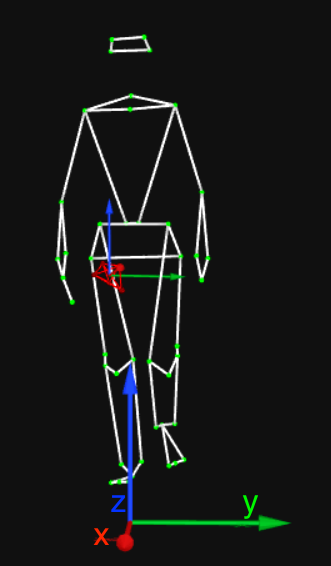}
\caption{Coordinate reference of the motion capture system, and the smartphone and MCU attached to the leg being tracked as a rigid body.}
\label{fig:coords_reference}
\end{figure}

An InvenSense MPU-9250~\cite{mpu2016} six-axis IMU mounted to an ESP8266 microcontroller (MCU) was used to capture triple-axis accelerometer data. A firmware to the ESP8266 was developed using Arduino Core libraries, to read raw accelerations by I2C protocol at a sampling rate of 100Hz. The acceleration measurements were read in $g$ units and transformed into units of $m/s^2$ using $g = 9.81m/s^2$. The MCU board was connected to the WiFi during its initialization. Then the accelerations were read from the MPU-9250, and sent using Open Sound Control (OSC) packages over UDP, following the defined rate. The smartphone and the MCU, screwed within a projected box, were attached to the leg using a band, as showed in  \autoref{fig:nexus_imu_band}. Five reflexive markers were taped in the band to calibrate it as a rigid body in the motion capture system, then at least three markers were kept during the walking trials to track it as a six-degree freedom object. These markers are also described in \autoref{tab:markers_details}, at the fifth first lines. In the \autoref{fig:coords_reference} is showed the motion capture software visualization of the coordinate reference calibrated for the optical motion capture system, smartphone and MCU attached to the leg being tracked as a rigid body. The MCU and smartphone were positioned to correspond to the motion capture reference in a way in which their coordinate systems are aligned.

An integration API was designed and developed to receive data from the MCU over UDP protocol and from the optical motion capture system by a real-time SDK provided by the Qualisys corporation~\cite{qtm2021realtime}. The three-axis accelerations and the 3D trajectories were acquired independently by this integration API but synchronously. The API starts to listen to the OSC packages from the MCU when the optical system begins the data acquisition. Also, the API stops listening to the OSC packages when the optical system finishes the data acquisition. Once the integration API guarantees both acquisitions initiate and finish together, and both systems have the same sampling rate, hence their data streams are synchronized.

\subsection*{Data acquisition} \label{subsec:data_acquisition}
Two data acquisition sessions were performed for each participant, and each one lasted about one hour. The sessions were performed on different days. In each session, the following procedure was adopted:

\begin{description}
\item[1) Calibration of the systems:] the optical motion capture system was calibrated by the Wand calibration method following the manufacture's instructions~\cite{qtm2011}. During the calibration, the coordinate system was defined as: \textit{x} was the direction in which the participant walked; \textit{y} was orthogonal to \textit{x}; and \textit{z} was orthogonal to both, pointing to the participants' head. The MCU also was turned on at this moment, and its communication to the computer was verified. As well as, the Android application was started to read and store the accelerometer readings in a CSV file. 

\item[2) Participant preparation:] the investigator showed the laboratory to the participant, explained the recording procedure, and asked the participant to sign the consent form. Further, the participant changed their clothes to tight-fitting outfits and wore tight caps to cover their hair. The investigator attached the reflexive markers presented in \autoref{fig:markerset} on the participants' skin or the tight-fitting clothes, using a proper double-sided tape. Also, the smartphone and the MCU box placed in the band were attached to the participant's leg. 

\item[3) Trials:] The participant was asked to stand up at the beginning of the walkway for few seconds to guarantee the proper function and communication of the systems. After these few seconds, the participant was asked to tap three times on the smartphone and MCU box (using their index finger, on which an additional marker was placed). Finally, the participant walked forth on a 5-m straight level walkway at their normal pace. At the end of the walkway, the participant stopped and tapped on the smartphone and MCU box again. The accelerations peaks generated by these taps allow a later verification of the data synchronization. Five trials of the participant walking using the band on their left leg were recorded, and five other sessions were recorded using the band placed on their right leg.

\item[4) Session ending:] After recording the ten trials, the band, and all the markers were removed from the participants' bodies. The Android application also was stopped.  The obtained visualizations from the session recordings were shown, and the scheduling of the participant's next session was confirmed.

\end{description}


\subsection*{Data processing}
The marker trajectories were labeled using the QTM software\footnote{QTM version 2018.1 (build 4220), RT Protocol versions 1.0 - 1.18 supported.}, as well as their gap-filling procedure. The investigator manually selected the best fit of interpolation in the trajectory editor~\cite{qtm2021processing} for each missing trajectory. Polynomial interpolation was applied in gaps smaller than ten frames, and relational interpolation, which is based on the movement of surrounding markers, was adopted for more complex cases (e.g., occlusions caused by the alignment of both legs during mid-stance and mid-swing phases). After that, the trajectories were smoothed, when necessary, using QTM software tool~\cite{qtm2021processing} by selecting a range of the trajectories and employing the appropriated filter. Small ranges of noisy data were locally smoothed using a moving average filter,  which is ideal to smooth spikes without affecting the  movements. Larger ranges of data containing expressive high-frequency noise were smoothed using a Butterworth low pass filter with a 5 Hz cut-off frequency. After that, these filled and smoothed trajectories were exported to the c3d\footnote{\url{https://www.c3d.org}} and Matlab file formats. These trajectories were then imported and processed using MoCap Toolbox\cite{BurgerToiviainen2013} under Matlab~\cite{MATLAB:2019b}. The exported trajectories were structured as \textit{MoCap data} from MoCap Toolbox, and processed to extract the temporal section containing only walking data, i.e., removing the beginning and end of the trials. The raw accelerations recorded by the MCU and smartphone were exported into CSV files, including the full trials, i.e., the beginning, walking data, and end of the trials. Additionally, the corresponding walking sections were extracted from these accelerations and exported to a CSV file. The walking sections obtained from the trajectories, MCU accelerations, and smartphone accelerations were assured to present temporal alignment.

\section*{Data Records}

All data are available from figshare~\cite{santos2021dataset}. They are organized in two folders: \textbf{raw\_data} containing the complete trials; and \textbf{processed\_data} storing the walking sections extracted from the raw data by removing the beginning and end of the trials. In both folders, the subjects' trial files are organized in sub folders associated to each subject identification: \textit{userID}, which IDs are from 01 to 30, not necessarily consecutive. A total of 20 trials have been recorded of each subject, 10 in \textit{day1} and 10 in \textit{day2}.

\subsection*{Raw data}

This folder stores the complete trials, including the beginning and end of them. These trials consists in 3D trajectories exported from the QTM software into c3d files, and MCU's accelerometer readings stored in CSV files. The c3d files are referenced as \textbf{capture\_userID\_dayD\_TT\_qtm.c3d}, and CSV files as \textbf{capture\_userID\_dayD\_TT\_imu.csv}, in which:

\begin{itemize}
    \item \textbf{userID} is the subject identification, as explained before;
    \item \textbf{dayD}  refers to the first or second day of the subject trials, \textbf{day1} or \textbf{day2};
    \item \textbf{TT} is the trial number, i.e. 0001 to 0010, for each \textbf{dayD}.
\end{itemize}

The description of the c3d labels are presented in \autoref{tab:markers_details}. All trajectories have the dimension $n \times 3$ (\textit{n} is the number of frames recorded), and the number format is real. Information about the columns of CSV files are presented in \autoref{tab:csv_details}. The corresponding frames to the accelerometer measurements are indicated in the first column of CSV files. In some cases, the frame counting of the accelerations does not start in one, but they are surely synchronized to the marker trajectories' frames.

\subsection*{Processed data}

This folder stores the walking sections obtained from the complete trials, removing the beginning and end of them. The walking sections of 3D trajectories were stored in Matlab files whereas the corresponding sections of accelerometer readings from the MCU and smartphone were stored in CSV files. The Matlab files are referenced as \textbf{capture\_userID\_dayD\_TT\_qtm\_walk.mat}, and CSV files as  \textbf{capture\_userID\_dayD\_TT\_imu\_walk.csv} and \textbf{capture\_userID\_dayD\_TT\_nexus\_walk.csv}, in which:

\begin{itemize}
    \item \textbf{userID} is the subject identification, as explained before;
    \item \textbf{dayD}  refers to the first or second day of the subject trials, \textbf{day1} or \textbf{day2};
    \item \textbf{TT} is the trial number, i.e. 0001 to 0010, for each \textbf{dayD}.
\end{itemize}

First, the c3d files containing the markers trajectories were read over Matlab, and their data were stored in a structure array from MoCap Toolbox named \textit{MoCap data}. The scheme and fields of this structure are described in \autoref{tab:mocap_details}. Format and set values of the fields are also presented. Some fields are composed by Structures, which are described on the  columns to the right. Those fields kept as default values of \textit{Mocap data} Structure are presented as empty or zero-valued in this table. Through the functions provided by Mocap Toolbox, walking sections were extracted by analysing the trajectories on \textit{x}-axis of feet markers, medial and lateral malleolus of both sides (details in \autoref{tab:markers_details}), to find the frame in which the subject displacement begins and ends. These frame numbers are stored in the field \textbf{frame\_init} and \textbf{frame\_end} of the Structure \textbf{other} of \textit{Mocap data}. 

Once the accelerometer readings were synchronized to the trajectories by the corresponding frame numbers, these yielded beginning and ending frames were also used to section the MCU accelerations.  The columns and scheme of CSV files which store the walking sections extracted from MCU accelerations are the same of the presented in \autoref{tab:csv_details}. The first value of column named \textbf{frame} presents the same amount of \textbf{frame\_init}, and the last value presents the same amount of \textbf{frame\_end} from the \textit{MoCap data} Structure.
 
 Accelerations from the smartphone were not capture synchronously to the marker trajectories, and only one CSV file stored all the accelerometer measurements of the 10 trials performed by a subject in each day. This CSV file was split into walking sections by correlation analysis between it and the walking sections of MCU accelerations, which were synchronized to the trajectories. Thus, these extracted walking sections from the smartphone accelerations do not contain the information about frame numbers although these were motion-aligned to the MCU's accelerometer readings. The columns of the CSV files that store the walking sections of the smartphone accelerations are described in \autoref{tab:csv_nexus_details}.

\section*{Technical Validation}

The presented multi-sensor dataset consists of three sources: an optical motion capture system, an IMU mounted to an ESP8266 board and an Android-based smartphone. The same experienced investigator attached the markers and the smartphone and MCU within a box mounted to a band on the participant's body for consistency purposes. 

\begin{description}
    \item[Optical motion capture system:] The calibration of the optical motion capture system was performed as described in the \textit{Data acquisition} of the Section \textit{Methods}, right before the beginning of each data acquisition session. In all achieved calibrations, the average residuals of each camera remained below 3 mm and similar among the 18 cameras. Also, the standard deviation between the actual wand length and the length perceived by the cameras was up to 1mm in the adopted calibrations. The gaps in the 3D trajectories of the markers were filled (i.e., all the trajectories are 100\% tracked), and the average residuals of 3D measured points fell below 4mm.
    
    \item[MCU:] The firmware was implemented using  Arduino Core libraries without adopting IMU libraries to avoid potentially applying of library default filters to the accelerations. Commonly, a scale range between $\pm2g$ and $\pm8g$ is adopted to the accelerometer in the gait literature~\cite{sprager2015inertial}. The scale range of $\pm2g$ is enough to measure the maximums instantaneous accelerations of normal walking (bellow to $1g$), and can capture more gait details (a scale range of $\pm2g$ corresponds to 16384 $LSB/g$ according to the MPU-9250 specification~\cite{mpu2016}). Thus, the firmware reads the registers' position of the MPU-9250 accelerometer and calculates the accelerations according to this sensors' sensitivity.
    
    \item[Smartphone:] The developed Android application uses the sensor routines provided by the platform~\cite{motion2021android} and the Android Open Source Project (AOSP)~\cite{aosp2021}. The same scale range of $\pm2g$ used to the MPU-9250 accelerometer was set to the MPU-6515 one using the Android Sensor API. This scale also corresponds to a specificity of 16384 $LSB/g$ for the MPU-6515. The Android API reads these accelerations from the IMU and calculates them in units of $m/s^2$, including the gravity.
\end{description}

\section*{Usage Notes}

The c3d files can be read by several free and open source tools which provide support to c3d files, such as EZC3D~\cite{michaudBiorbd2021} and Motion Kinematic \& Kinetic Analyzer (MOKKA)~\cite{mokka2021}. The Matlab files store the structure \textit{MoCap data} from MoCap Toolbox\cite{BurgerToiviainen2013}, and can be manipulated using the functions provided by this Toolbox. It presents several routines to visualize, perform kinematics and kinetics analysis and apply projections on the data. This Toolbox supports any kind of marker set. The CSV format files can be read using any text or spreadsheet editor, as well as by common functions over Matlab\footnote{\url{https://www.mathworks.com/help/matlab/ref/readtimetable.html}} or Python\footnote{\url{https://pandas.pydata.org/pandas-docs/stable/reference/api/pandas.read_csv.html}}.

\section*{Code availability}

The developed Matlab and Python codes to process the data are freely available on the first author's github repository\footnote{\url{https://github.com/geisekss/motion_capture_analysis}}. The MoCap Toolbox is freely available and extensively documented on the University of Jyväskylä website\footnote{\url{https://www.jyu.fi/hytk/fi/laitokset/mutku/en/research/materials/mocaptoolbox}}.

\bibliography{main}

\begin{thebibliography}{10}
\urlstyle{rm}
\expandafter\ifx\csname url\endcsname\relax
  \def\url#1{\texttt{#1}}\fi
\expandafter\ifx\csname urlprefix\endcsname\relax\def\urlprefix{URL }\fi
\expandafter\ifx\csname doiprefix\endcsname\relax\def\doiprefix{DOI: }\fi
\providecommand{\bibinfo}[2]{#2}
\providecommand{\eprint}[2][]{\url{#2}}

\bibitem{baker2007history}
\bibinfo{author}{Baker, R.}
\newblock \bibinfo{journal}{\bibinfo{title}{The history of gait analysis before
  the advent of modern computers}}.
\newblock {\emph{\JournalTitle{Gait \& posture}}}
  \textbf{\bibinfo{volume}{26}}, \bibinfo{pages}{331--342}
  (\bibinfo{year}{2007}).

\bibitem{muro2014gait}
\bibinfo{author}{Muro-De-La-Herran, A.}, \bibinfo{author}{Garcia-Zapirain, B.}
  \& \bibinfo{author}{Mendez-Zorrilla, A.}
\newblock \bibinfo{journal}{\bibinfo{title}{Gait analysis methods: An overview
  of wearable and non-wearable systems, highlighting clinical applications}}.
\newblock {\emph{\JournalTitle{Sensors}}} \textbf{\bibinfo{volume}{14}},
  \bibinfo{pages}{3362--3394} (\bibinfo{year}{2014}).

\bibitem{gouwanda2008emerging}
\bibinfo{author}{Gouwanda, D.} \& \bibinfo{author}{Senanayake, S.}
\newblock \bibinfo{title}{Emerging trends of body-mounted sensors in sports and
  human gait analysis}.
\newblock In \emph{\bibinfo{booktitle}{4th Kuala Lumpur Internation`al
  Conference on Biomedical Engineering}}, \bibinfo{pages}{715--718}
  (\bibinfo{organization}{Springer}, \bibinfo{year}{2008}).

\bibitem{winter2009biomechanics}
\bibinfo{author}{Winter, D.}
\newblock \emph{\bibinfo{title}{Biomechanics and motor control of human
  movement}}, vol. \bibinfo{volume}{4th edition} (\bibinfo{publisher}{John
  Wiley \& Sons}, \bibinfo{year}{2009}).

\bibitem{morris2010review}
\bibinfo{author}{Morris, R.} \& \bibinfo{author}{Lawson, S.}
\newblock \bibinfo{journal}{\bibinfo{title}{A review and evaluation of
  available gait analysis technologies, and their potential for the measurement
  of impact transmission}}.
\newblock {\emph{\JournalTitle{Newcastle University}}}  (\bibinfo{year}{2010}).

\bibitem{ramirez2017review}
\bibinfo{author}{Ramirez-Bautista, J.}, \bibinfo{author}{Huerta-Ruelas, J.},
  \bibinfo{author}{Chaparro-C{\'a}rdenas, S.} \&
  \bibinfo{author}{Hern{\'a}ndez-Zavala, A.}
\newblock \bibinfo{journal}{\bibinfo{title}{A review in detection and
  monitoring gait disorders using in-shoe plantar measurement systems}}.
\newblock {\emph{\JournalTitle{IEEE reviews in biomedical engineering}}}
  \textbf{\bibinfo{volume}{10}}, \bibinfo{pages}{299--309}
  (\bibinfo{year}{2017}).

\bibitem{yang2019review}
\bibinfo{author}{Yang, G.}, \bibinfo{author}{Tan, W.}, \bibinfo{author}{Jin,
  H.}, \bibinfo{author}{Zhao, T.} \& \bibinfo{author}{Tu, L.}
\newblock \bibinfo{journal}{\bibinfo{title}{Review wearable sensing system for
  gait recognition}}.
\newblock {\emph{\JournalTitle{Cluster Computing}}}
  \textbf{\bibinfo{volume}{22}}, \bibinfo{pages}{3021--3029}
  (\bibinfo{year}{2019}).

\bibitem{kaufman1999future}
\bibinfo{author}{Kaufman, K.}
\newblock \bibinfo{journal}{\bibinfo{title}{Future directions in gait
  analysis}}.
\newblock {\emph{\JournalTitle{Gait analysis in the science of
  rehabilitation}}} \bibinfo{pages}{85--112} (\bibinfo{year}{1998}).

\bibitem{akhtaruzzaman2016gait}
\bibinfo{author}{Akhtaruzzaman, M.}, \bibinfo{author}{Shafie, A.} \&
  \bibinfo{author}{Khan, M.~R.}
\newblock \bibinfo{journal}{\bibinfo{title}{Gait analysis: Systems,
  technologies, and importance}}.
\newblock {\emph{\JournalTitle{Journal of Mechanics in Medicine and Biology}}}
  \textbf{\bibinfo{volume}{16}}, \bibinfo{pages}{1630003}
  (\bibinfo{year}{2016}).

\bibitem{mayagoitia2002accelerometer}
\bibinfo{author}{Mayagoitia, R.}, \bibinfo{author}{Nene, A.} \&
  \bibinfo{author}{Veltink, P.}
\newblock \bibinfo{journal}{\bibinfo{title}{Accelerometer and rate gyroscope
  measurement of kinematics: an inexpensive alternative to optical motion
  analysis systems}}.
\newblock {\emph{\JournalTitle{Journal of biomechanics}}}
  \textbf{\bibinfo{volume}{35}}, \bibinfo{pages}{537--542}
  (\bibinfo{year}{2002}).

\bibitem{lee2010identifying}
\bibinfo{author}{Lee, J.~B.}, \bibinfo{author}{Sutter, K.~J.},
  \bibinfo{author}{Askew, C.~D.} \& \bibinfo{author}{Burkett, B.~J.}
\newblock \bibinfo{journal}{\bibinfo{title}{Identifying symmetry in running
  gait using a single inertial sensor}}.
\newblock {\emph{\JournalTitle{Journal of Science and Medicine in Sport}}}
  \textbf{\bibinfo{volume}{13}}, \bibinfo{pages}{559--563}
  (\bibinfo{year}{2010}).

\bibitem{moore2015elaborate}
\bibinfo{author}{Moore, J.}, \bibinfo{author}{Hnat, S.} \&
  \bibinfo{author}{van~den Bogert, A.}
\newblock \bibinfo{journal}{\bibinfo{title}{An elaborate data set on human gait
  and the effect of mechanical perturbations}}.
\newblock {\emph{\JournalTitle{PeerJ}}} \textbf{\bibinfo{volume}{3}},
  \bibinfo{pages}{e918} (\bibinfo{year}{2015}).

\bibitem{fukuchi2018public}
\bibinfo{author}{Fukuchi, C.}, \bibinfo{author}{Fukuchi, R.} \&
  \bibinfo{author}{Duarte, M.}
\newblock \bibinfo{journal}{\bibinfo{title}{A public dataset of overground and
  treadmill walking kinematics and kinetics in healthy individuals}}.
\newblock {\emph{\JournalTitle{PeerJ}}} \textbf{\bibinfo{volume}{6}},
  \bibinfo{pages}{e4640} (\bibinfo{year}{2018}).

\bibitem{schreiber2019multimodal}
\bibinfo{author}{Schreiber, C.} \& \bibinfo{author}{Moissenet, F.}
\newblock \bibinfo{journal}{\bibinfo{title}{A multimodal dataset of human gait
  at different walking speeds established on injury-free adult participants}}.
\newblock {\emph{\JournalTitle{Scientific data}}} \textbf{\bibinfo{volume}{6}},
  \bibinfo{pages}{1--7} (\bibinfo{year}{2019}).

\bibitem{ngo2014largest}
\bibinfo{author}{Ngo, T.}, \bibinfo{author}{Makihara, Y.},
  \bibinfo{author}{Nagahara, H.}, \bibinfo{author}{Mukaigawa, Y.} \&
  \bibinfo{author}{Yagi, Y.}
\newblock \bibinfo{journal}{\bibinfo{title}{The largest inertial sensor-based
  gait database and performance evaluation of gait-based personal
  authentication}}.
\newblock {\emph{\JournalTitle{Pattern Recognition}}}
  \textbf{\bibinfo{volume}{47}}, \bibinfo{pages}{228--237}
  (\bibinfo{year}{2014}).

\bibitem{gadaleta2018idnet}
\bibinfo{author}{Gadaleta, M.} \& \bibinfo{author}{Rossi, M.}
\newblock \bibinfo{journal}{\bibinfo{title}{Idnet: Smartphone-based gait
  recognition with convolutional neural networks}}.
\newblock {\emph{\JournalTitle{Pattern Recognition}}}
  \textbf{\bibinfo{volume}{74}}, \bibinfo{pages}{25--37}
  (\bibinfo{year}{2018}).

\bibitem{luo2020database}
\bibinfo{author}{Luo, Y.} \emph{et~al.}
\newblock \bibinfo{journal}{\bibinfo{title}{A database of human gait
  performance on irregular and uneven surfaces collected by wearable sensors}}.
\newblock {\emph{\JournalTitle{Scientific data}}} \textbf{\bibinfo{volume}{7}},
  \bibinfo{pages}{1--9} (\bibinfo{year}{2020}).

\bibitem{santos2020manifold}
\bibinfo{author}{Santos, G.} \emph{et~al.}
\newblock \bibinfo{journal}{\bibinfo{title}{Manifold learning for user
  profiling and identity verification using motion sensors}}.
\newblock {\emph{\JournalTitle{Pattern Recognition}}}
  \textbf{\bibinfo{volume}{106}}, \bibinfo{pages}{107408}
  (\bibinfo{year}{2020}).

\bibitem{freire2020evaluation}
\bibinfo{author}{Freire, S.}, \bibinfo{author}{Santos, G.},
  \bibinfo{author}{Armondes, A.}, \bibinfo{author}{Meneses, E.} \&
  \bibinfo{author}{Wanderley, M.}
\newblock \bibinfo{journal}{\bibinfo{title}{Evaluation of inertial sensor data
  by a comparison with optical motion capture data of guitar strumming
  gestures}}.
\newblock {\emph{\JournalTitle{Sensors}}} \textbf{\bibinfo{volume}{20}},
  \bibinfo{pages}{5722} (\bibinfo{year}{2020}).

\bibitem{lapinski2019wide}
\bibinfo{author}{Lapinski, M.} \emph{et~al.}
\newblock \bibinfo{journal}{\bibinfo{title}{A wide-range, wireless wearable
  inertial motion sensing system for capturing fast athletic biomechanics in
  overhead pitching}}.
\newblock {\emph{\JournalTitle{Sensors}}} \textbf{\bibinfo{volume}{19}},
  \bibinfo{pages}{3637} (\bibinfo{year}{2019}).

\bibitem{qualisys2021}
\bibinfo{author}{Qualisys}.
\newblock \bibinfo{title}{Qualisys motion capture systems}
  (\bibinfo{year}{accessed on June of 2021}).

\bibitem{leardini2007new}
\bibinfo{author}{Leardini, A.} \emph{et~al.}
\newblock \bibinfo{journal}{\bibinfo{title}{A new anatomically based protocol
  for gait analysis in children}}.
\newblock {\emph{\JournalTitle{Gait \& posture}}}
  \textbf{\bibinfo{volume}{26}}, \bibinfo{pages}{560--571}
  (\bibinfo{year}{2007}).

\bibitem{davis1991gait}
\bibinfo{author}{Davis~III, R.}, \bibinfo{author}{Ounpuu, S.},
  \bibinfo{author}{Tyburski, D.} \& \bibinfo{author}{Gage, J.}
\newblock \bibinfo{journal}{\bibinfo{title}{A gait analysis data collection and
  reduction technique}}.
\newblock {\emph{\JournalTitle{Human movement science}}}
  \textbf{\bibinfo{volume}{10}}, \bibinfo{pages}{575--587}
  (\bibinfo{year}{1991}).

\bibitem{qtm2011}
\bibinfo{author}{Qualisys}.
\newblock \emph{\bibinfo{title}{Qualisys track manager: Getting Started}}.
\newblock \bibinfo{address}{Gothenburg, Sweden} (\bibinfo{year}{accessed on
  June of 2021}).

\bibitem{mpu2016}
\bibinfo{author}{InvenSense}.
\newblock \emph{\bibinfo{title}{MPU-9250 Product Specification Revision 1.1}}.
\newblock \bibinfo{address}{California, USA} (\bibinfo{year}{2016}).

\bibitem{qtm2021realtime}
\bibinfo{author}{Qualisys}.
\newblock \emph{\bibinfo{title}{QTM Real-time Server Protocol Documentation
  Version 1.22}}.
\newblock \bibinfo{address}{Gothenburg, Sweden} (\bibinfo{year}{Accessed on
  June of 2021}).

\bibitem{qtm2021processing}
\bibinfo{author}{Qualisys}.
\newblock \emph{\bibinfo{title}{Qualisys track manager: Processing your data}}.
\newblock \bibinfo{address}{Gothenburg, Sweden} (\bibinfo{year}{accessed on
  June of 2021}).

\bibitem{BurgerToiviainen2013}
\bibinfo{author}{Burger, B.} \& \bibinfo{author}{Toiviainen, P.}
\newblock \bibinfo{title}{{MoCap Toolbox -- A Matlab toolbox for computational
  analysis of movement data}}.
\newblock In \bibinfo{editor}{Bresin, R.} (ed.)
  \emph{\bibinfo{booktitle}{Proceedings of the 10th Sound and Music Computing
  Conference}}, \bibinfo{pages}{172--178} (\bibinfo{publisher}{KTH Royal
  Institute of Technology}, \bibinfo{address}{Stockholm, Sweden},
  \bibinfo{year}{2013}).

\bibitem{MATLAB:2019b}
\bibinfo{organization}{The Mathworks, Inc.}, \bibinfo{address}{Massachusetts,
  USA}.
\newblock \emph{\bibinfo{title}{{MATLAB version 9.7.0.1296695 (R2019b)}}}
  (\bibinfo{year}{2019}).

\bibitem{santos2021dataset}
\bibinfo{author}{Santos, G.}, \bibinfo{author}{Wanderley, M.},
  \bibinfo{author}{Tavares, T.} \& \bibinfo{author}{Rocha, A.}
\newblock \bibinfo{title}{A multi-sensor and cross-session human gait dataset
  captured through an optical system and inertial measurement units},
  \url{10.6084/m9.figshare.14727231} (\bibinfo{year}{2021}).

\bibitem{sprager2015inertial}
\bibinfo{author}{Sprager, S.} \& \bibinfo{author}{Juric, M.}
\newblock \bibinfo{journal}{\bibinfo{title}{Inertial sensor-based gait
  recognition: A review}}.
\newblock {\emph{\JournalTitle{Sensors}}} \textbf{\bibinfo{volume}{15}},
  \bibinfo{pages}{22089--22127} (\bibinfo{year}{2015}).

\bibitem{motion2021android}
\bibinfo{author}{Developers, A.}
\newblock \emph{\bibinfo{title}{Motion sensors}} (\bibinfo{year}{Accessed on
  June of 2021}).

\bibitem{aosp2021}
\bibinfo{author}{Project, A. O.~S.}
\newblock \emph{\bibinfo{title}{Sensors}} (\bibinfo{year}{Accessed on June of
  2021}).

\bibitem{michaudBiorbd2021}
\bibinfo{author}{Michaud, B.} \& \bibinfo{author}{Begon, M.}
\newblock \bibinfo{journal}{\bibinfo{title}{ezc3d: An easy c3d file i/o
  cross-platform solution for c++, python and matlab}}.
\newblock {\emph{\JournalTitle{Journal of Open Source Software}}}
  \textbf{\bibinfo{volume}{6}}, \bibinfo{pages}{2911} (\bibinfo{year}{2021}).

\bibitem{mokka2021}
\bibinfo{author}{Barré, A.}
\newblock \bibinfo{title}{Mokka: Motion kinematic and kinetic analyzer}
  (\bibinfo{year}{Accessed on June of 2021}).

\end{thebibliography}

\section*{Acknowledgements} 

We thank the financial support of CAPES DeepEyes project (\# 88882.160443/2014-01) and FAPESP DéjàVu (\# 2017/12646-3). We also thank to CIRMMT staff and members for the support during this study, also all volunteers for their participation.

\section*{Author contributions statement}

G.S. designed the protocol, performed the data acquisition, and processed the data. M.W. provided the technical support and resources to establish the data collection. M.W. and G.S discussed the experimental design. G.S. T.T. and A.R. contributed to the data analysis and the manuscript writing. All authors reviewed the manuscript.

\section*{Competing interests}

The authors declare no competing interests.

\newpage

\section*{Figures \& Tables}

\newpage

\begin{longtable}{|l|l|p{10cm}|}
\caption{Description and placement of each reflexive marker attached to the participants' body.} \label{tab:markers_details} \\
    \hline
    \textbf{Labels} & \textbf{Description}                 & \textbf{Landmarks for marker placement}                                                                                     \\ \toprule \bottomrule
    top\_right      & Marker on the top right of the band  & Located approximately on the top right of the smartphone                                                      \\ \hline
    top\_left       & Marker on the top left of the band   & Located approximately on the top left of the smartphone                                                       \\ \hline
    top\_imu        & Marker on the center of the band     & Located approximately on the top of the MCU's box                                                             \\ \hline
    head\_imu       & Marker on the left half of the band  & Located approximately on the left side of the MCU's box                                                       \\ \hline
    power           & Marker on the right half of the band & Located approximately on top of the powerbank used as power source to the MCU                                 \\ \hline
    finger          & Index finger                         & Placed on the distal phalanx of the right/left index finger according to the leg which the band was placed on \\ \hline
    head\_FR        & Right front head                     & Located approximately over the right temple                                                                   \\ \hline
    head\_FL        & Left front head                      & Located approximately over the left temple                                                                    \\ \hline
    head\_BR        & Right back head                      & Placed on the back of the head, roughly in a horizontal line of the right front head marker                   \\ \hline
    head\_BL        & Left left head                       & Placed on the back of the head, roughly in a horizontal line of the left front head marker                    \\ \hline
    shoulder\_R     & Right shoulder                       & Placed on the acromio-clavicular joint                                                                        \\ \hline
    shoulder\_L     & Left shoulder                        & Placed on the acromio-clavicular joint                                                                        \\ \hline
    c7              & 7th Cervical Vertebrae               & Spinous process of the 7th cervical vertebrae                                                                 \\ \hline
    clav            & Clavicule                            & Jugular Notch where the clavicles meet the sternum                                                            \\ \hline
    elbow\_R        & Right elbow                          & Lateral epicondyle approximating elbow joint axis                                                             \\ \hline
    elbow\_L        & Left elbow                           & Lateral epicondyle approximating elbow joint axis                                                             \\ \hline
    wrist\_BR       & Right wrist marker B                 & Right wrist bar pinkie side                                                                                   \\ \hline
    wrist\_AR       & Right wrist marker A                 & Right wrist bar thumb side                                                                                    \\ \hline
    wrist\_AL       & Left wrist marker A                  & Left wrist bar thumb side                                                                                     \\ \hline
    wrist\_BL       & Left wrist marker B                  & Left wrist bar pinkie side                                                                                    \\ \hline
    finger\_R       & Right finger                         & Dorsum of the right hand just below the head of the second metacarpal                                         \\ \hline
    finger\_L       & Left finger                          & Dorsum of the left hand just below the head of the second metacarpal                                          \\ \hline
    ips\_R          & Left PSIS                            & Right posterior-superior iliac spine                                                                          \\ \hline
    ips\_L          & Left PSIS                            & Left posterior-superior iliac spine                                                                           \\ \hline
    ias\_R          & Right ASIS                           & Right anterior-superior iliac spine                                                                           \\ \hline
    ias\_L          & Left ASIS                            & Left anterior-superior iliac spine                                                                            \\ \hline
    ftc\_R          & Right greater trochanter             & Most lateral prominence of the right greater trochanter                                                       \\ \hline
    ftc\_L          & Left greater trochanter              & Most lateral prominence of the left greater trochanter                                                        \\ \hline
    fme\_R          & Left medial femoral epicondyle       & Most medial prominence of the left medial femoral epicondyle                                                  \\ \hline
    fme\_L          & Right medial femoral epicondyle      & Most medial prominence of the right medial femoral epicondyle                                                 \\ \hline
    tcc\_R          & Right tibial tuberosity              & Most anterior border of the right tibial tuberosity                                                           \\ \hline
    ttc\_L          & Left tibial tuberosity               & Most anterior border of the left tibial tuberosity                                                            \\ \hline
    fax\_R          & Right fibula head                    & Proximal tip of the head of the right fibula                                                                  \\ \hline
    fax\_L          & Left fibula head                     & Proximal tip of the head of the left fibula                                                                   \\ \hline
    fle\_R          & Right lateral femoral epicondyle     & Most lateral prominence of the right lateral femoral epicondyle                                               \\ \hline
    fle\_L          & Left lateral femoral epicondyle      & Most lateral prominence of the left lateral femoral epicondyle                                                \\ \hline
    heel\_R         & Right heel                           & Right posterior calcaneus                                                                                     \\ \hline
    heel\_L         & Left heel                            & Left posterior calcaneus                                                                                      \\ \hline
    fal\_R          & Right lateral malleolus              & Lateral prominence of the right lateral tibial malleolus                                                      \\ \hline
    fal\_L          & Left lateral malleolus               & Lateral prominence of the left lateral tibial malleolus                                                       \\ \hline
    tam\_R          & Right medial malleolus               & Most medial prominence of the right medial tibial malleolus                                                   \\ \hline
    tam\_L          & Left medial malleolus                & Most medial prominence of the left medial tibial malleolus                                                    \\ \hline
    fm5\_R          & Right 5th metatarsal head            & Dorsal margin of the right fifth metatarsal head                                                              \\ \hline
    fm5\_L          & Left 5th metatarsal head             & Dorsal margin of the left fifth metatarsal head                                                               \\ \hline
    fm2\_R          & Right 2nd metatarsal head            & Dorsal aspect of the right second metatarsal head                                                             \\ \hline
    fm2\_L          & Left 2nd metatarsal head             & Dorsal aspect of the left second metatarsal head                                                              \\ \hline
    fm1\_R          & Right 1st metatarsal head            & Dorsal margin of the right first metatarsal head                                                              \\ \hline
    fm1\_L          & Left 1st metatarsal head             & Dorsal margin of the left first metatarsal head                                                               \\ \hline

\end{longtable}

\begin{table}[ht]
\centering
\caption{\label{tab:csv_details} Description of the columns of the CSV files generated by the MCU accelerations during the subjects' trials.}
\begingroup
    \renewcommand{\arraystretch}{1.2}
    \begin{tabular}{|l|l|l|}
        \hline
        \textbf{Columns} & \textbf{Description} & \textbf{Format} \\ \toprule \bottomrule
        \textbf{Frame} & Frame number to correspond to the marker trajectories' frames & Integer  \\ \hline
        \textbf{Acc x} & x-axis of accelerometer reading in the referenced frame number & Real (in units of $m/s^2$) \\ \hline
        \textbf{Acc y} & y-axis of accelerometer reading in the referenced frame number & Real (in units of $m/s^2$) \\ \hline
        \textbf{Acc z} & z-axis of accelerometer reading in the referenced frame number & Real (in units of $m/s^2$) \\ \hline
 \end{tabular}
\endgroup
\end{table}

\begin{table}[ht]
\centering
\caption{\label{tab:csv_nexus_details} Description of the columns of the CSV files generated by the smartphone accelerations during the subjects' trials.}
\begingroup
    \renewcommand{\arraystretch}{1.2}
    \begin{tabular}{|l|p{8cm}|l|}
        \hline
        \textbf{Columns} & \textbf{Description} & \textbf{Format} \\ \toprule \bottomrule
        \textbf{Acc x} & x-axis of the accelerometer measurement & Real (in units of $m/s^2$) \\ \hline
        \textbf{Acc y} & y-axis of the accelerometer measurement & Real (in units of $m/s^2$) \\ \hline
        \textbf{Acc z} & z-axis of the accelerometer measurement & Real (in units of $m/s^2$) \\ \hline
 \end{tabular}
\endgroup
\end{table}   

\begin{table}[ht]
\begin{center}
    \caption{\label{tab:mocap_details} Description of the scheme, fields, their format and set values on \textit{Mocap data} structure.}
    \begingroup
    \renewcommand{\arraystretch}{1.2}
        \begin{tabular}{|l|l|l|l|l|p{5cm}|}
        \hline
        \textbf{Field name}   & \textbf{Format}   & \multicolumn{4}{l|}{\textbf{Value}}                                                                         \\ \hline
        \textbf{type}         & String            & \multicolumn{4}{l|}{`MoCap data'}                                                                           \\ \hline
        \textbf{filename}     & String            & \multicolumn{4}{l|}{}                                                                                       \\ \hline
        \textbf{nFrames}      & Integer           & \multicolumn{4}{l|}{\textit{f}}                                                                                      \\ \hline
        \textbf{nCameras}     & Integer           & \multicolumn{4}{l|}{18}                                                                                     \\ \hline
        \textbf{nMarkers}     & Integer           & \multicolumn{4}{l|}{47}                                                                                     \\ \hline
        \textbf{freq}         & Integer           & \multicolumn{4}{l|}{100}                                                                                    \\ \hline
        \textbf{nAnalog}      & Integer           & \multicolumn{4}{l|}{0}                                                                                      \\ \hline
        \textbf{anaFreq}      & Integer           & \multicolumn{4}{l|}{0}                                                                                      \\ \hline
        \textbf{timederOrder} & Integer           & \multicolumn{4}{l|}{0}                                                                                      \\ \hline
        \textbf{markerName}   & String              & \multicolumn{4}{l|}{47 x 1 String}                                                                           \\ \hline
        \textbf{data}         & Real              & \multicolumn{4}{l|}{\textit{f} x 141 Real (in units of mm)}                                                                          \\ \hline
        \textbf{analogdata}   & Real              & \multicolumn{4}{l|}{}                                                                                       \\ \hline
        \multirow{11}{*}{\textbf{other}} & \multirow{11}{*}{Structure} & \textbf{descr}                        & \multicolumn{3}{l|}{String}                                         \\ \cline{3-6} 
                                 &                             & \textbf{timeStamp}                    & \multicolumn{3}{l|}{String}                                         \\ \cline{3-6} 
                                 &                             & \textbf{dataIncluded}                 & \multicolumn{3}{l|}{`3D'}                                           \\ \cline{3-6} 
                                 &                             & \multirow{6}{*}{\textbf{RigidBodies}} & \multirow{6}{*}{Structure} & \textbf{Bodies}    & 1                 \\ \cline{5-6} 
                                 &                             &                                       &                            & \textbf{Name}      & `imu\_nexus\_box' \\ \cline{5-6} 
                                 &                             &                                       &                            & \textbf{Positions} & 1 x 3 x f Real (in units of mm)   \\ \cline{5-6} 
                                 &                             &                                       &                            & \textbf{Rotations} & 1 x 9 x f Real  (rotation matrices)  \\ \cline{5-6} 
                                 &                             &                                       &                            & \textbf{RPYs}      & 1 x 3 x f Real (in units of degrees -- roll, pitch and yaw)  \\ \cline{5-6} 
                                 &                             &                                       &                            & \textbf{Residual}  & 1 x 1 x f Real (in units of mm)  \\ \cline{3-6} 
                                 &                             & \textbf{frame\_init}                  & \multicolumn{3}{l|}{Integer}                                        \\ \cline{3-6} 
                                 &                             & \textbf{frame\_end}                   & \multicolumn{3}{l|}{Integer}  \\ \hline
        \end{tabular}
    \endgroup
    \vskip 1ex
    \begin{minipage}{0.75\linewidth}
    \textsuperscript{*}f is the number of frames of the walking section.
    
    All fields' names are in bold. The ones which format is Structure, their fields are detailed on the rightmost columns.
    The  empty or zero-valued fields were not filled in the \textit{Mocap data} structure. 
    \end{minipage}
\end{center}
\end{table}   
    
\end{document}